\documentclass[12pt]{iopart}
\usepackage{iopams}  
\usepackage{graphicx}
\usepackage{multirow}
\usepackage[table]{xcolor}
\usepackage{cite}
\expandafter\let\csname equation*\endcsname\relax
\expandafter\let\csname endequation*\endcsname\relax
\usepackage{amsmath}
\begin{document}

\title{The Relativity Principle at the Foundation of Quantum Mechanics}

\author{W.M. Stuckey$^1$, Timothy McDevitt$^2$, and Michael Silberstein$^{3,4}$}

\address{$^1$ Department of Physics, Elizabethtown College, Elizabethtown, PA 17022, USA}
\address{$^2$ Department of Mathematical Sciences, Elizabethtown College, Elizabethtown, PA 17022, USA}
\address{$^3$ Department of Philosophy, Elizabethtown College, Elizabethtown, PA 17022, USA}
\address{$^4$ Department of Philosophy, University of Maryland, College Park, MD 20742, USA}

\ead{stuckeym@etown.edu}

\begin{abstract}
Quantum information theorists have created axiomatic reconstructions of quantum mechanics (QM) that are very successful at identifying precisely what distinguishes quantum probability theory from classical and more general probability theories in terms of information-theoretic principles. Herein, we show how one such principle, Information Invariance \& Continuity, at the foundation of those axiomatic reconstructions maps to ``no preferred reference frame'' (NPRF, aka ``the relativity principle'') as it pertains to the invariant measurement of Planck's constant $h$ for Stern-Gerlach (SG) spin measurements. This is in exact analogy to the relativity principle as it pertains to the invariant measurement of the speed of light $c$ at the foundation of special relativity (SR). Essentially, quantum information theorists have extended Einstein's use of NPRF from the boost invariance of measurements of $c$ to include the SO(3) invariance of measurements of $h$ between different reference frames of mutually complementary spin measurements via the principle of Information Invariance \& Continuity. Consequently, the ``mystery'' of the Bell states that is responsible for the Tsirelson bound and the exclusion of the no-signalling, ``superquantum'' Popescu-Rohrlich joint probabilities is understood to result from conservation per Information Invariance \& Continuity between different reference frames of mutually complementary qubit measurements, and this maps to conservation per NPRF in spacetime. If one falsely conflates the relativity principle with the classical theory of SR, then it may seem impossible that the relativity principle resides at the foundation of non-relativisitic QM. In fact, there is nothing inherently classical or quantum about NPRF. Thus, the axiomatic reconstructions of QM have succeeded in producing a principle account of QM that reveals as much about Nature as the postulates of SR. 
\end{abstract}
\vspace{2pc}

\newpage

\section{Introduction}
Feynman famously said, ``I think I can safely say that nobody understands quantum mechanics'' \cite{feynman1964}. Despite the fact that quantum mechanics ``has survived all tests'' and ``we all know how to use it and apply it to problems,'' Gell-Mann agreed with Feynman saying, ``we have learned to live with the fact that nobody can understand it'' \cite{gellmann1993}. As a result, there are many programs designed to \textit{interpret} quantum mechanics (QM), i.e., reveal what QM is telling us about Nature. We will not review such attempts here (the interested reader is referred to Drummond's 2019 overview of QM interpretations \cite{drummond2019}), rather in this paper we will explain how axiomatic reconstructions of QM based on information-theoretic principles (e.g., see \cite{hardy2001,fuchs2002,galindo2002,barrett2007,bruknerZeil2009,dakicBrukner2009,masanesMuller2011,chiribellaDarianoPerinotti2011,hardy2011,hardy2016,goyal2010,kochen2015,oeckl2016,hohn2018,hohnWever2017,masanesMullerAugPerez2013,torreMasanesShortMuller2012,fivel2012,barnumMullerUdudec2014,koberinski2018} or the review by Jaeger \cite{jaeger2018}) contain a surprising advance in the understanding of QM. Specifically, we will show how the principle of Information Invariance \& Continuity \cite{bruknerZeil2009}:
\begin{quote}
    The total information of one bit is invariant under a continuous change between different complete sets of mutually complementary measurements.
\end{quote}
at the basis of information-theoretic reconstructions of QM already implies the relativity principle (aka ``no preferred reference frame (NPRF)'') as it pertains to the invariant measurement of Planck's constant $h$ when applied to spin-$\frac{1}{2}$ measurements in spacetime. This is in total analogy to the Lorentz transformations of special relativity (SR) being based on the relativity principle as it pertains to the invariant measurement of the speed of light $c$ (light postulate). Thus, the information-theoretic reconstructions of QM (hereafter, ``reconstructions of QM'') provide a ``principle'' account of QM in total analogy to that of SR \cite{Bub1999,vancamp2011,koberinski2018,felline2018}, revealing a deep unity between these pillars of modern physics where others have perceived tension \cite{bellbook,mamone,alford2016,PR1994}.

Before proceeding further, some caveats are in order. First, the relativity principle, i.e., ``The laws of physics must be the same in all inertial reference frames'' or NPRF for short, is not restricted to ``The laws of classical physics,'' it applies to all of physics. Second, that it resides at the foundation of a theory does not mean the theory is ``relativistic.'' For example, NPRF is at the foundation of Newtonian mechanics with its Galilean transformations and Newtonian mechanics is certainly ``non-relativistic.'' Thus, there is no reason a priori to exclude the possibility that NPRF resides at the foundation of non-relativistic QM. 

And, our use of NPRF deals exclusively with the kinematic structure underlying QM, i.e., denumerable-dimensional Hilbert space (of arbitrarily large, but finite, dimension). This is in total analogy to the relativity principle underwriting the the kinematic structure of SR, i.e., Minkowski spacetime (M4). In both cases the kinematic structure constrains but does not dictate the dynamics. Bub writes \cite{bub2020}:
\begin{quote}
The information-theoretic interpretation is the proposal to take Hilbert space as the kinematic framework for the physics of an indeterministic universe, just as Minkowski space provides the kinematic framework for the physics of a non-Newtonian, relativistic universe. In special relativity, the geometry of Minkowski space imposes spatio-temporal constraints on events to which the relativistic dynamics is required to conform. In quantum mechanics, the non-Boolean projective geometry of Hilbert space imposes objective kinematic (i.e., pre-dynamic) probabilistic constraints on correlations between events to which a quantum dynamics of matter and fields is required to conform.
\end{quote}
For example, the relativity principle is responsible for the light postulate and together they give M4 at the foundation of SR, but M4 does not dictate the contents of the 4-momentum vector. Likewise, reconstructions of QM give the qubit Hilbert space structure at the foundation of QM, but that qubit Hilbert space structure does not dictate the Hamiltonian for the propagator. 

Finally, we should also point out that our result is not related to quantum reference frames \cite{bruknerReframes,bruknerReframes2,bruknerReframes3}, Lorentz invariance of entangled states \cite{lamata2006}, the relativity principle in QM per Davis \cite{davis1977} or Dragan \& Ekert \cite{Dragan2020}, or the relativity of simultaneity applied to quantum experiments \cite{RoS2017,arraut2021}. The relativity principle will be applied herein to reference frames related by spatial rotations SO(3). More specifically, these spatial reference frames will be those established by mutually complementary qubit measurements \cite{bruknerZeil2009} per the ``closeness requirement'' between quantum theory and classical measurement devices  \cite{dakic2013,dakicgroup} (Section \ref{SecQubit}). While spatial rotations plus Lorentz boosts constitute the restricted Lorentz group, spatial rotations plus Galilean boosts constitute the homogeneous Galilean group, so the use of the relativity principle here does not imply Lorentz invariance. Again, conformity to NPRF does not mean a theory is ``relativistic'' in that sense. 

The relativity principle has a long history in physics, e.g., Galileo used the relativity of motion principle to argue against geocentricism and Newtonian mechanics is invariant under Galilean transformations. Einstein generalized Galileo's version of the relativity principle, ``The laws of mechanics must be the same in all inertial reference frames'' to ``The laws of physics must be the same in all inertial reference frames,'' so he could apply it to the value of $c$ from Maxwell's equations \cite{norton2004,serway,moylan2021}. As Norton points out, Maxwell's discovery of $c$ plus NPRF makes M4 of SR inevitable \cite{norton2004}. Here we will see that Planck's discovery of $h$ plus NPRF makes the denumerable-dimensional Hilbert space of QM inevitable as well.  

Quantum information theorists engaged in reconstructions of QM have stated specifically their desire to discover principles at the foundation of QM analogous to NPRF and light postulate at the foundation of SR  \cite{fuchs1,chiribella1,hardy2016,dakicBrukner2009,mullergroup,barnumMullerUdudec2014,kochen2015,masanesMullerAugPerez2013}. If not the relativity principle specifically, at least principles that can ``be translated back into language of physics'' \cite{chiribellaDarianoPerinotti2011}. Of course, different reconstructions of QM contain different information-theoretic principles precisely because quantum information scientists ``design algorithms and protocols at an abstract level, without considering whether they will be implemented with light, atoms or any other type of physical substrate'' \cite{masanesMullerAugPerez2013}. Nonetheless, they all reveal directly or indirectly that the key difference between classical and quantum probability theories resides in the continuity of reversible transformations between pure states (Section \ref{SecQubit}). In what is considered the first axiomatic reconstruction of QM \cite{jaeger2018}, Hardy notes that by adding the single word ``continuous'' to his reversibility axiom one obtains quantum probability theory instead of classical probability theory \cite{hardy2001}. Indeed, many authors emphasize this point \cite{hardy2011,masanesMullerAugPerez2013,barnumMullerUdudec2014,dakicgroup}, e.g., Koberinski \& M\"uller write \cite{koberinski2018}:
\begin{quote}
    We suggest that (continuous) reversibility may be the postulate which comes closest to being a candidate for a glimpse on the genuinely physical kernel of ``quantum reality''. Even though Fuchs may want to set a higher threshold for a ``glimpse of quantum reality'', this postulate is quite surprising from the point of view of classical physics: when we have a discrete system that can be in a finite number of perfectly distinguishable alternatives, then one would classically expect that reversible evolution must be discrete too. For example, a single bit can only ever be flipped, which is a discrete indivisible operation. Not so in quantum theory: the state $|0\rangle$ of a qubit can be continuously-reversibly ``moved over'' to the state $|1\rangle$. For people without knowledge of quantum theory (but of classical information theory), this may appear as surprising or ``paradoxical'' as Einstein's light postulate sounds to people without knowledge of relativity. 
\end{quote}
Our goal here is to show how this key difference between classical and quantum probability theories per the principle of Information Invariance \& Continuity relates directly to an application of NPRF in spacetime. 

Of course, as a ``principle'' account of QM the information-theoretic reconstructions do not provide ``a constructive account of ontological structure'' that many deem necessary for a \textit{true} interpretation of QM \cite{koberinski2018,brownTimp2006}. Einstein noted the difference between ``principle'' and ``constructive'' theories in this famous passage \cite{einstein1919}:
\begin{quote}
   We can distinguish various kinds of theories in physics. Most of them are constructive. They attempt to build up a picture of the more complex phenomena out of the materials of a relatively simple formal scheme from which they start out. [The kinetic theory of gases is an example.] ... Along with this most important class of theories there exists a second, which I will call ``principle-theories.'' These employ the analytic, not the synthetic, method. The elements which form their basis and starting point are not hypothetically constructed but empirically discovered ones, general characteristics of natural processes, principles that give rise to mathematically formulated criteria which the separate processes or the theoretical representations of them have to satisfy. [Thermodynamics is an example.] ... The advantages of the constructive theory are completeness, adaptability, and clearness, those of the principle theory are logical perfection and security of the foundations. The theory of relativity belongs to the latter class.
\end{quote}
Nearly every introductory physics textbook introduces SR via the relativity principle and light postulate without qualifying that introduction as somehow lacking an ``interpretation.'' With few exceptions, physicists have come to accept the principles of SR without worrying about a constructive counterpart. Thus, a principle account of QM based on NPRF as with SR certainly constitutes an important advance in our understanding of QM. Perhaps prophetically, Bell said \cite[p. 85]{bell1997}:
\begin{quote}
 I think the problems and puzzles we are dealing with here will be cleared up, and ... our descendants will look back on us with the same kind of superiority as we now are tempted to feel when we look at people in the late nineteenth century who worried about the aether. And Michelson-Morley ..., the puzzles seemed insoluble to them. And came Einstein in nineteen five, and now every schoolboy learns it and feels ... superior to those old guys. Now, it's my feeling that all this action at a distance and no action at a distance business will go the same way. But someone will come up with the answer, with a reasonable way of looking at these things. If we are lucky it will be to some big new development like the theory of relativity. 
\end{quote}
By revealing the relativity principle's role at the foundation of QM, information-theoretic reconstructions of QM have revealed what QM is telling us about Nature to no less an extent than SR. And, SR's principle explanation of Nature certainly constituted a ``big new development'' for physics in 1905. 
As emphasized by Fuchs, ``Where present-day quantum-foundation studies have stagnated in the stream of history is not so unlike where the physics of length contraction and time dilation stood before Einstein's 1905 paper on special relativity'' \cite{fuchs2002}. At that time, ``Maxwellian physicists were ready to abandon the relativity of motion principle'' \cite{moylan2021} and even ``Einstein was willing to sacrifice the greatest success of 19th century physics, Maxwell’s theory, seeking to replace it by one conforming to an emission theory of light, as the classical, Galilean kinematics demanded'' before realizing that such an emission theory would not work \cite{norton2004}. Thus, concerning his decision to produce a principle explanation instead of a constructive explanation for time dilation and length contraction, Einstein writes \cite{einstein1949}:
\begin{quote}
   By and by I despaired of the possibility of discovering the true laws by means of constructive efforts based on known facts. The longer and the more despairingly I tried, the more I came to the conviction that only the discovery of a universal formal principle could lead us to assured results.
\end{quote}
Therefore, being in a similar situation today with QM, it is not unreasonable to seek a compelling principle account of QM along the lines of SR. Again, a principle account of QM that maps to NPRF applied to $h$ at its foundation would be as valuable to understanding QM as NPRF applied to $c$ is to understanding SR and, as we will show, the information-theoretic reconstructions of QM entail exactly that understanding. 

We start in Section \ref{SecQubit} with an introduction to the relevant information-theoretic formalism on the qubit at the basis of the reconstructions of QM. This introduction is not a mathematically detailed exposition on the reconstructions of QM (for that see \cite{mullerlectures} and as related to this paper \cite{Manko2021}). Rather, in this section we introduce only the key information-theoretic concepts associated with the qubit in the reconstructions of QM, as required to make our argument to the physicist interested in foundations of QM (but not necessarily familiar with quantum information theory). Virtually all undergraduate physics textbooks introduce the counterintuitive concepts of time dilation, length contraction, and the relativity of simultaneity using the relativity principle and the invariant measurement of $c$ at the foundation of SR. Our goal here is to present an equally accessible introduction to the counterintuitive concept of the qubit using the relativity principle and the invariant measurement of $h$ at the foundation of QM, as implied by the reconstructions of QM. Here we will see that the information-theoretic principles of Existence of an Information Unit and Continuous Reversibility \cite{masanesMullerAugPerez2013}, or in combined form Information Invariance \& Continuity, already reveal a role for NPRF at the foundation of QM.

In Section \ref{SecSpin}, we finish our argument by looking at the role played by Planck's constant $h$ in QM and its relation to the Existence of an Information Unit. In particular, we focus on three facts: QM obtains because $h \ne 0$, $h$ is a universal constant of Nature, and Stern-Gerlach (SG) spin measurements constitute the invariant measurement of $h$. We then show why continuous reversibility in SG spin measurements of $h$ is ``quite surprising from the point of view of classical physics'' \cite{koberinski2018}, i.e., there is no constructive classical model for it and it leads to ``average-only'' projection of spin angular momentum. Most generally, Information Invariance \& Continuity leads to ``average-only'' projection/transmission/... between the different reference frames of mutually complementary qubit measurements \cite{bruknerZeil2009,bruknerZeil2003}. In Section \ref{SecBell}, we review how this result leads to the counterintuitive ``average-only'' conservation characterizing quantum entanglement per the Bell states \cite{stuckey2019,stuckey2020,silberstein2021}. Thus, ``average-only'' conservation responsible for the Tsirelson bound is explained by conservation per Information Invariance \& Continuity (conservation per NPRF in spacetime). In Section \ref{SecPRbox}, we show how conservation per the relativity principle can be used to rule out the no-signalling, ``superquantum'' joint probabilities of Popescu \& Rohrlich \cite{PR1994} in spacetime. We conclude in Section \ref{SecConcl}.

\section{\label{SecQubit}The Qubit and the Relativity Principle}

We start by noting the term ``quantum state'' can refer to the probability amplitude vector, e.g., $|u\rangle$, or to the probability (density) matrix $\rho$. It will be clear which is meant by the context. Next, we review the difference between the classical bit and the qubit per Hardy \cite{hardy2001}, starting with the qubit.

In a 2-dimensional (2D) Hilbert space spanned by $|u\rangle$ and $|d\rangle$, a general state $|\psi \rangle$ is given by $|\psi \rangle = c_1|u\rangle + c_2|d\rangle$ with $c_1$ and $c_2$ complex and $|c_1|^2 + |c_2|^2 = 1$. In general, such 2D states are called qubits and the density matrix is given by $\rho = |\psi\rangle \langle \psi|$. In quantum information theory, these qubits represent an elementary piece of information for quantum systems; a quantum system is probed and one of two possible outcomes obtains, e.g., yes/no, up/down, pass/no pass, etc. The structure of all such binary systems from an information-theoretic perspective is identical. 

\newpage

\begin{figure}
\begin{center}
\includegraphics [height = 65mm]{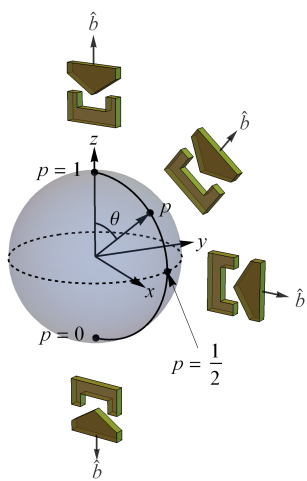}
\caption{Probability state space for the qubit $|u\rangle$ in the $z$ basis. Since this state space is isomorphic to 3-dimensional real space, the Bloch sphere is shown in a real space reference frame with its related Stern-Gerlach (SG) magnet orientations (see Knight \cite[p. 1307]{knight} for an explanation of the SG experiment). The probability is given for a $+1$ outcome at the measurement direction shown \cite{bruknerZeil2003}. Compare this with Figure \ref{SGExp2}.} \label{Qubit}
\end{center}
\end{figure}

\begin{figure}
\begin{center}
\includegraphics [height = 50mm]{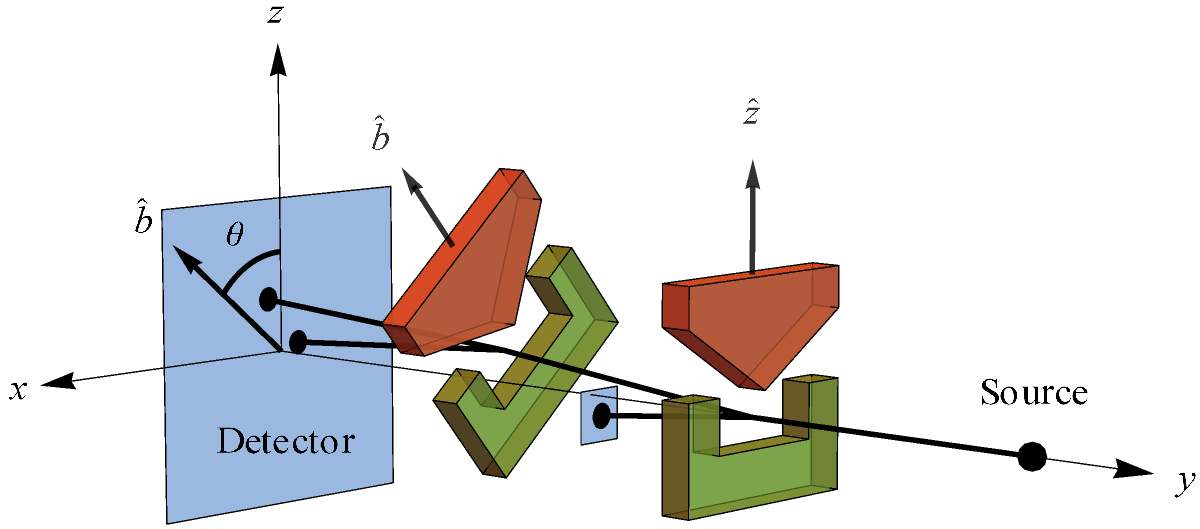}  
\caption{In this set up, the first SG magnets (oriented at $\hat{z}$) are being used to produce an initial state $|\psi\rangle = |u\rangle$ for measurement by the second SG magnets (oriented at $\hat{b}$). Compare this with Figure \ref{Qubit}. } \label{SGExp2}
\end{center}
\end{figure}

A general Hermitian measurement operator in 2D Hilbert space has outcomes given by its (real) eigenvalues and can be written $(\lambda_1)|1\rangle\langle 1| + (\lambda_2)|2\rangle\langle 2|$ where $|1\rangle$ and $|2\rangle$ are the eigenstates for the eigenvalues $\lambda_1$ and $\lambda_2$, respectively. Any such Hermitian matrix $M$ can be expanded in the Pauli matrices
$$
\sigma_x = \left ( \begin{array}{rr} 0 & \phantom{00}1 \\ 1 & \phantom{0}0 \end{array} \right ), \quad
\sigma_y = \left ( \begin{array}{rr} 0 & \phantom{0}-\textbf{i} \\ \textbf{i} & 0  \end{array} \right ), \quad \mbox{and} \quad
\sigma_z =\left ( \begin{array}{rr} 1 & 0 \\ 0 & \phantom{0}-1 \end{array} \right )
$$
plus the identity matrix $\textbf{I}$
\begin{equation}
 M = m_0 \textbf{I} +  m_x\sigma_x + m_y\sigma_y + m_z\sigma_z 
\end{equation}
where $(m_0,m_x,m_y,m_z)$ are real. The eigenvalues of $M$ are given by
$$
m_0 \pm \sqrt{m_x^2 + m_y^2 + m_z^2}.
$$
We see that $m_x, m_y, m_z$ give two eigenvalues centered about $m_0$. All measurement operators with the same eigenvalues are related by SU(2) transformations given by some combination of $e^{\textbf{i}\Theta\sigma_j}$, where $j = \{x,y,z\}$ and $\Theta$ is an angle in Hilbert space. Any density matrix can be expanded in the same fashion
\begin{equation}
 \rho = \frac{1}{2}\left(I +  \rho_x\sigma_x + \rho_y\sigma_y + \rho_z\sigma_z \right)
\end{equation}
where $(\rho_x,\rho_y,\rho_z)$ are real. The Bloch sphere is defined by $\rho_x^2 + \rho_y^2 + \rho_z^2 = 1$ with pure states residing on the surface of the sphere and mixed states residing inside the sphere, consistent with the fact that mixed states contain less information than pure states. Since we will be referring to spin-$\frac{1}{2}$ measurements and states later, we will denote our eigenstates $|u\rangle$ for spin up with eigenvalue $+1$ and $|d\rangle$ for spin down with eigenvalue $-1$. [See Figures \ref{Qubit} and \ref{SGExp2} associated with the state space corresponding to $|u\rangle$ in the $\sigma_z$ basis.] The transformations relating various pure states on the sphere are continuously reversible so that in going from a pure state to a pure state one always passes through other pure states. This is distinctly different from the reversibility axiom between pure states for classical probability theory's fundamental unit of information, the classical bit, as noted above above by Koberinski \& M\"uller \cite{koberinski2018}. 

\begin{figure}
\begin{center}
\includegraphics [height = 55mm]{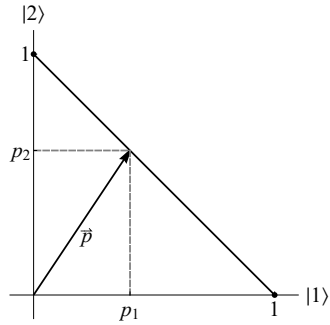}  \caption{Probability state space for the classical bit.} \label{ClassBit}
\end{center}
\end{figure}

In classical probability theory, the only continuous way to get from one pure discrete state for a classical bit to the other pure state is through mixed states. For example, suppose we place a single ball in one of two boxes labeled 1 and 2 with probabilities $p_1$ and $p_2$, respectively. The probability state is given by the vector $\vec{p} = \left ( \begin{array}{c} p_1 \\ p_2\end{array} \right )$. When normalized we have $\vec{p} = \left ( \begin{array}{c} p_1 \\ 1 - p_1\end{array} \right )$ which can be represented by a line segment in the plane connecting $\left ( \begin{array}{c} 1 \\ 0\end{array} \right )$ and $\left ( \begin{array}{c} 0 \\ 1\end{array} \right )$ (Figure \ref{ClassBit}). We see here that reversibility of pure states is discrete, i.e., accomplished via permutation \cite{hardy2001}.

Notice that the state space for the classical bit is 1-dimensional and represented by a $2 \times 1$ matrix while that for the qubit is 3-dimensional and represented by a $2\times 2$ matrix. In general, the dimension of the probability space for the generalized bit (gbit) of a general probability theory is $d = 2^s - 1$ with $s = 1, 2, 3, ... $ and the gbit is represented by a $2 \times 2 \times \ldots \times 2$ tensor ($d$ equals the number of 2's) \cite{paterekDakBruk2010}.  Having seen the fundamental difference between classical probability theory and quantum probability theory per their fundamental units of information, we now review how higher-dimensional Hermitian operators in Hilbert space are related to the qubit.

\newpage

The structure of the qubit is important because any higher-dimensional Hermitian matrix with the same eigenvalues can be constructed via SU(2) and the qubit from the diagonal version, as explained by Hardy \cite{hardy2001}. For example, suppose you want to construct the $L_x$ measurement operator with eigenvalues $+1,0,-1$ in the $L_z$ eigenbasis $|u\rangle = \left ( \begin{array}{rrr} 1 \\ 0 \\ 0\end{array} \right )$, $|0\rangle = \left ( \begin{array}{rrr} 0 \\ 1 \\ 0\end{array} \right )$, $|d\rangle = \left ( \begin{array}{rrr} 0 \\ 0 \\ 1\end{array} \right )$. You simply use the SU(2) transformation $e^{\textbf{i}\Theta\sigma_x}$ sequentially, first about the $|d\rangle$ axis with $\Theta = 90^{\circ}$, then about the post transformed $|u\rangle$ axis with $\Theta = 45^{\circ}$, finally about the post transformed $|0\rangle$ axis with $\Theta = -45^{\circ}$. That generates $\displaystyle L_x = \frac{1}{\sqrt{2}}\left ( \begin{array}{rrr} 0 & \phantom{00}1 & \phantom{00} 0 \\ 1 & \phantom{0}0 & 1 \\0 & 1 & 0 \end{array} \right )$ from $\displaystyle L_z = \left ( \begin{array}{rrr} 1 & \phantom{00}0 & \phantom{00} 0 \\ 0 & \phantom{0}0 & 0 \\0 & 0 & -1 \end{array} \right )$. To obtain the third member of the mutually complementary measurements, $L_y$, simply use the SU(2) transformation $e^{\textbf{i}\Theta\sigma_x}$ about the $|d\rangle$ axis with $\Theta = -90^{\circ}$, then again use the SU(2) transformation $e^{\textbf{i}\Theta\sigma_x}$ about the post transformed $|u\rangle$ axis with $\Theta = 45^{\circ}$, finally use the SU(2) transformation $e^{\textbf{i}\Theta\sigma_y}$ about the post transformed $|0\rangle$ axis with $\Theta = 45^{\circ}$. That generates $\displaystyle L_y = \frac{1}{\sqrt{2}}\left ( \begin{array}{rrr} 0 & \phantom{0}-\textbf{i} & \phantom{000} 0 \\ \textbf{i} & \phantom{0}0 & -\textbf{i} \\0 & \textbf{i} & 0 \end{array} \right )$ from $L_z$ giving the algebra of mutually complementary measurements $[L_x,L_y] = iL_z$, cyclic. In this example, the Pauli matrices $(\sigma_x,\sigma_y,\sigma_z)$ are clearly visible in the matrices $(L_x,L_y,L_z)$, respectively. The completion of the reconstruction uses the tensor product $\otimes$ to add particles for any given dimension in accord with the ``Locality'' axiom \cite{dakicBrukner2009} and the relevant dynamical transformation of the state for the Schr\"odinger equation \cite{hardy2001}. [For more information on this mathematical structure and its extension to the probability states for a continuous random variable see \cite{Manko2021}.] But, we don't need to proceed further in the reconstruction process, as we have what we need to show the reconstruction of QM via Information Invariance \& Continuity maps to NPRF giving rise to the invariant measurement of $h$ in spacetime. 

\newpage

To summarize, we see that the reconstruction of QM builds the denumerable-dimensional Hilbert space structure of QM foundationally upon a fundamental 2D object, the qubit. The difference between classical probability and quantum probability then resides in the fact that a pure state can be transformed into another pure state through other pure states in continuous fashion for the qubit, while that is not possible for the classical bit. Of course, this continuously reversible transformation property would also hold for a gbit with $s > 2$, so why does Nature prefer the qubit? 

In order to argue for the qubit from the gbit, Masanes et al. employed ``Continuous Reversibility, Tomographic Locality and Existence of an Information Unit, No Simultaneous Encoding, All Effects Are Observable, and Gbits Can Interact'' \cite{masanesMullerAugPerez2013,mullergroup}. In other words, arguing for the qubit while keeping to the very general information-theoretic principles is highly non-trivial. Dakic \& Brukner presented an argument based on their ``closeness requirement: the dynamics of a single elementary system can be generated by the invariant interaction between the system and a `macroscopic transformation device' that is itself described within the theory in the macroscopic (classical) limit'' \cite{dakic2013,dakicgroup}. This is due to the fact that the measuring devices used to measure quantum systems are themselves made from quantum systems. For example, the classical magnetic field of an SG magnet is used to measure the spin of spin-$\frac{1}{2}$ particles and that classical magnetic field ``can be seen as a limit of a large coherent state, where a large number of spin-$\frac{1}{2}$ particles are all prepared in the same quantum state'' \cite{bruknergroup}. 

Per Brukner \& Zeilinger \cite{bruknerZeil2009,bruknerZeil2003}, if we identify the preparation state $|\psi\rangle = |u\rangle$ at $\hat{z}$ with the reference frame of mutually complementary spin measurements $[J_x,J_y,J_z]$ ($J_i = \frac{\hbar}{2}\sigma_i$), then the closeness requirement means our reference frame of mutually complementary measurements is $[\hat{x},\hat{y},\hat{z}]$ in real space. Thus, they depict the Bloch sphere in that real space reference frame with associated SG magnet orientations a la Figure \ref{Qubit} \cite{bruknerZeil2003}. Unless otherwise noted, we also make this association throughout, so that ``the reference frame of a complete set of mutually complementary measurements'' is simply ``the reference frame.'' While it may prove necessary to consider generalizations of QM that require higher-dimensional space in order to produce a theory of quantum gravity \cite{hardyQG2007,dakic2013}, the reconstruction of QM clearly shows that one may consider the qubit to reside at the foundation of the denumerable-dimensional Hilbert space structure of QM. And, as we will now see, the qubit structure already reveals a role for the relativity principle at the foundation of QM.

Again, by ``relativity principle'' we mean ``The laws of physics must be the same in all inertial reference frames,'' aka NPRF. In SR, we are concerned with the fact that everyone measures the same speed light $c$, regardless of their motion relative to the source (light postulate). Here the inertial reference frames are characterized by motion at constant velocity relative to the source and different reference frames are related by Lorentz boosts. Since $c$ is a constant of Nature per Maxwell's equations, NPRF implies the light postulate (or ``NPRF + $c$'' for short) \cite{serway,knight}. Thus, we see that NPRF + $c$ resides at the foundation of SR. 

Likewise, we have seen that different 2D Hilbert space measurement operators with the same outcomes are related by SU(2) transformations and that SU(2) transformations in Hilbert space map to SO(3) rotations between different reference frames in 3-dimensional real space (Information Invariance \& Continuity). In information-theoretic terms, the total knowledge one has about the elementary system must be independent of how they choose to represent that knowledge \cite{bruknerZeil2009}. Since spatial rotations, like Lorentz boosts, relate inertial reference frames, the information-theoretic qubit structure reveals a role for NPRF at the foundation of QM. 

Since we have so far only reviewed the very general structure of the qubit per information-theoretic principles, we don't as yet have a fundamental constant of Nature in play. However, we can already see that the qubit implies a role for the relativity principle (NPRF) in QM. To complete our analogy with SR and its NPRF + $c$, we need to relate all of this to the fundamental constant of Nature at the foundation of QM, i.e., Planck's constant $h$.

\section{\label{SecSpin}Planck's Constant and Spin}

Planck introduced $h$ in his explanation of blackbody radiation and we now understand that electromagnetic radiation with frequency $f$ is comprised of indivisible quanta (photons) of energy $hf$. One difference between the classical view of a continuous electromagnetic field and the quantum reality of photons is manifested in polarization measurements. According to classical electromagnetism, there is no non-zero lower limit to the energy of polarized electromagnetic radiation that can be transmitted by a polarizing filter. However, given that the radiation is actually composed of indivisible photons, there is a non-zero lower limit to the energy passed by a polarizing filter, i.e., each quantum of energy $hf$ either passes or it doesn't. Thus, we understand that the classical ``expectation'' of fractional amounts of quanta can only obtain on average per the quantum reality, so we expect the corresponding quantum theory will be probabilistic. In information-theoretic terms, a system is composed fundamentally of discrete units of finite information (the qubit). Since the qubit contains finite information, it cannot contain enough information to account for the outcomes of every possible measurement done on it. Thus, a theory of qubits must be probabilistic \cite{zeilinger1999,bruknergroup,bruknerZeil1999}. Of course, the relationship between classical and quantum mechanics per its expectation values is another textbook result, e.g., the Ehrenfest theorem. 

And, the fact that classical results are obtained from quantum results for $h \rightarrow 0$ is common knowledge. In information-theoretic terms, $h$ represents ``a universal limit on how much simultaneous information is accessible to an observer'' \cite{hohn2018}. For example, $[X,P] = \textbf{i}\hbar$ means there is a trade-off between what one can know simultaneously about the position and momentum of a quantum system. If $h = 0$, as in classical mechanics, there is no such limit to this simultaneous knowledge. For spin-$\frac{1}{2}$ measurements $J_i$, we have $[J_x,J_y]=\textbf{i}\hbar J_z$, cyclic. So we see that $h \ne 0$ in this case corresponds to the existence of a set of mutually complementary spin measurements associated with the reference frame shown in Figure \ref{Qubit} (more on this in Section \ref{SecConcl}). 

Given that $h$ is a constant of Nature, NPRF dictates that everyone measure the same value for it (``Planck postulate'' in analogy with the light postulate) and the measurement of spin via SG magnets constitutes a measurement of $h$ \cite{weinberg2017}. Again, the $\pm 1$ eigenvalues of the Pauli matrices correspond to $\pm \frac{\hbar}{2}$ for spin-$\frac{1}{2}$ measurement outcomes. Thus, the general result from Information Invariance \& Continuity concerning the SO(3) invariance of measurement outcomes for a qubit implies NPRF + $h$ (relativity principle $\rightarrow$ Planck postulate) for QM in total analogy to NPRF + $c$ (relativity principle $\rightarrow$ light postulate) for SR. One consequence of the continuously reversible movement of one qubit state to another when referring to an SG spin measurement is ``average-only'' projection. 

\begin{figure}
\begin{center}
\includegraphics [height = 75mm]{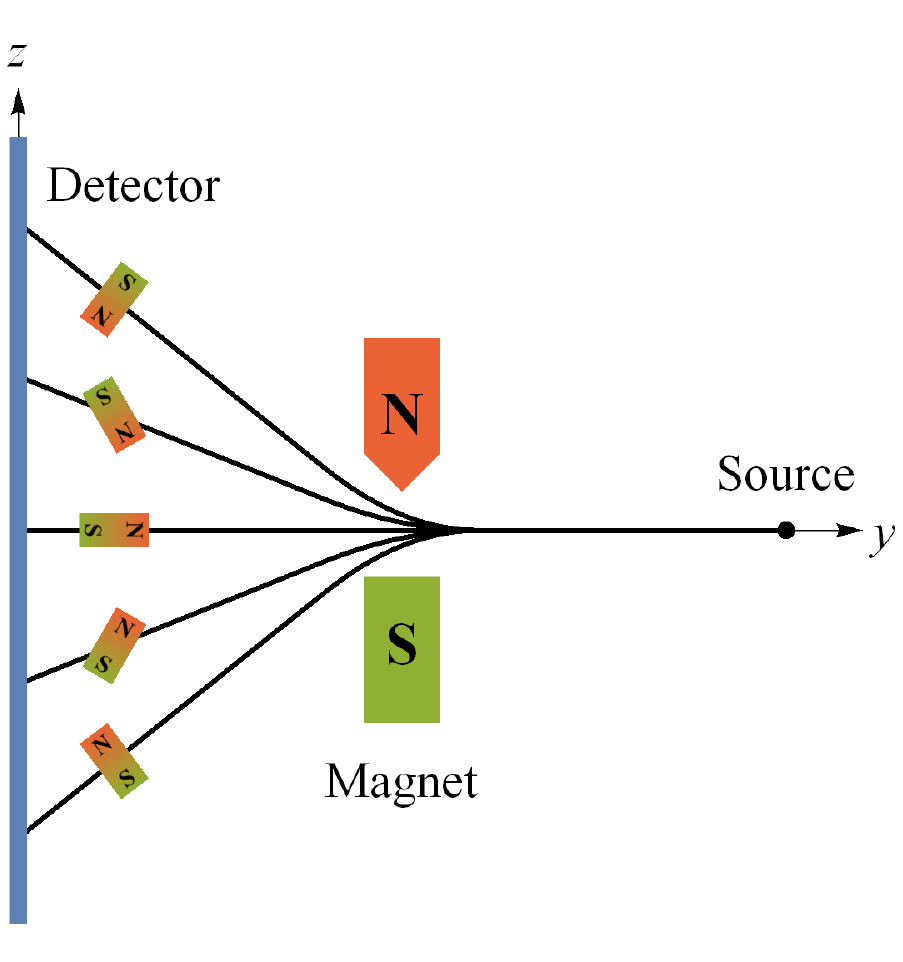}
\caption{\textbf{The classical constructive model of the Stern-Gerlach (SG) experiment.} If the atoms enter with random orientations of their ``intrinsic'' magnetic moments (due to their ``intrinsic'' angular momenta), the SG magnets should produce all possible deflections, not just the two that are observed \cite{knight,franklin2019}.} \label{SGclassical}
\end{center}
\end{figure}

\begin{figure}
\begin{center}
\includegraphics [height = 65mm]{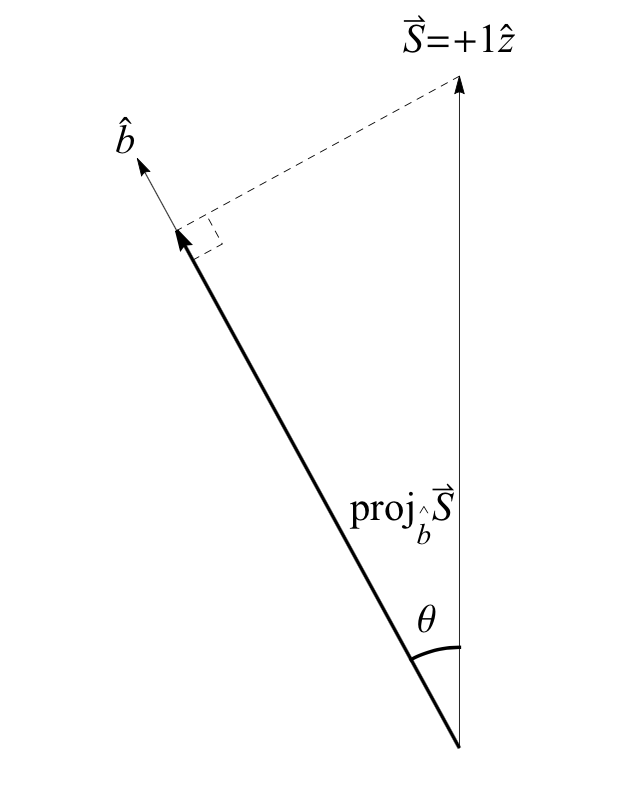} 
\caption{The ``intrinsic'' angular momentum of Bob's particle $\vec{S}$ projected along his measurement direction $\hat{b}$. This does \textit{not} happen with spin angular momentum due to NPRF.} \label{Projection}
\end{center}
\end{figure}

\begin{figure}
\begin{center}
\includegraphics [height = 75mm]{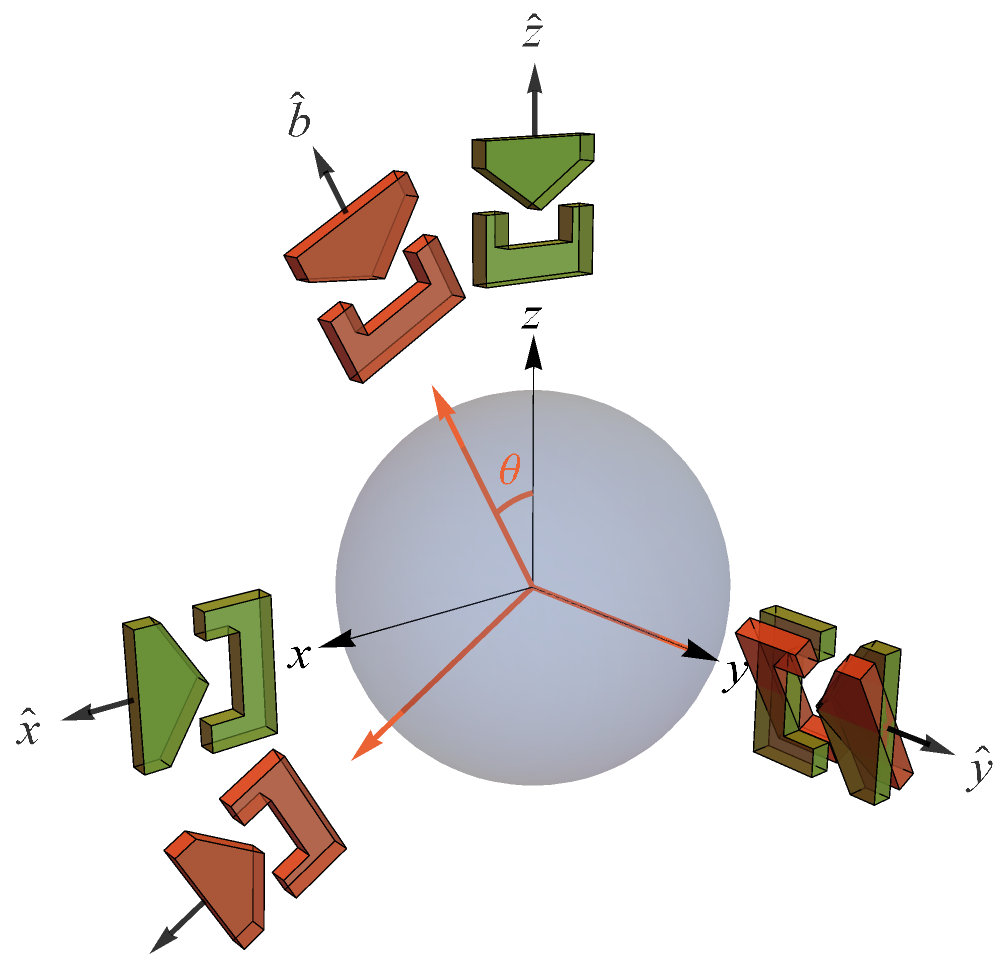} 
\caption{State space for a qubit showing two reference frames of mutually complementary SG spin measurements \cite{bruknerZeil2003}.} \label{ComplBases}
\end{center}
\end{figure}

Suppose we create a preparation state oriented along the positive $z$ axis as in Figure \ref{Qubit}, i.e., $|\psi\rangle = |u\rangle$, so that our ``intrinsic'' angular momentum is $\vec{S} = +1\hat{z}$ (in units of $\frac{\hbar}{2} = 1$). Now proceed to make a measurement with the SG magnets oriented at $\hat{b}$ making an angle $\theta$ with respect to $\hat{z}$ (Figure \ref{SGExp2}). According to the constructive account of classical physics \cite{knight,franklin2019} (Figure \ref{SGclassical}), we expect to measure $\vec{S}\cdot\hat{b} = \cos{(\theta)}$ (Figure \ref{Projection}), but we cannot measure anything other than $\pm 1$ due to NPRF (contra the prediction by classical physics). As a consequence, we can only recover $\cos{(\theta)}$ \textit{on average}, i.e., NPRF dictates ``average-only'' projection
\begin{equation}
(+1) P(+1 \mid \theta) + (-1) P(-1 \mid \theta) = \cos (\theta) \label{AvgProjection}
\end{equation}
Of course, this is precisely $\langle\sigma\rangle$ per QM. Eq. (\ref{AvgProjection}) with our normalization condition \linebreak $P(+1 \mid \theta) +  P(-1 \mid \theta) = 1$ then gives 
\begin{equation}
P(+1 \mid \theta) = \mbox{cos}^2 \left(\frac{\theta}{2} \right) \label{UPprobability}
\end{equation}
and 
\begin{equation}
P(-1 \mid \theta) = \mbox{sin}^2 \left(\frac{\theta}{2} \right) \label{DOWNprobability}
\end{equation} 
again, precisely in accord with QM. And, if we identify the preparation state $|\psi\rangle = |u\rangle$ at $\hat{z}$ with the reference frame of mutually complementary spin measurements $[J_x,J_y,J_z]$, then the SG measurement at $\hat{b}$ constitutes a reference frame of mutually complementary measurements rotated by $\theta$ in real space relative to the reference frame of the preparation state (Figure \ref{ComplBases}). Thus, ``average-only'' projection follows from Information Invariance \& Continuity when applied to SG measurements in real space.

The fact that one obtains $\pm 1$ outcomes at some SG magnet orientation is not mysterious per se, it can be accounted for by the classical constructive model in Figure \ref{SGclassical}. The constructive account of the $\pm 1$ outcomes would be one of particles with ``intrinsic'' angular momenta and therefore ``intrinsic'' magnetic moments \cite{knight} orientated in two opposite directions in space, parallel or anti-parallel to the magnetic field. Given this constructive account of the $\pm 1$ outcomes at this particular SG magnet orientation, we would then expect that the varying orientation of the SG magnetic field with respect to the magnetic moments, created as we rotate our SG magnets, would cause the degree of deflection to vary. Indeed, this is precisely the constructive account that led some physicists to expect all possible deflections for the particles as they passed through the SG magnets, having assumed that these particles would be entering the SG magnetic field with random orientations of their ``intrinsic'' magnetic moments \cite{franklin2019} (Figure \ref{SGclassical}). But according to this constructive account, if the $\pm 1$ outcomes constitute a measurement of $h$ in accord with the rest of quantum physics, then our rotated orientations would not be giving us the value for $h$ required by quantum physics otherwise. Indeed, a rotation of $90^\circ$ would yield absolutely no deflection at all (akin to measuring the speed of a light wave as zero when moving through the aether at speed $c$). That would mean our original SG magnet orientation would constitute a preferred frame in violation of the relativity principle, NPRF. Essentially, as Michelson and Morley rotated their interferometer the constructive model predicted they would see a change in the interference pattern \cite{michelson}, but instead they saw no change in the interference pattern in accord with NPRF. Likewise, as Stern and Gerlach rotated their magnets the constructive model predicted they would see a change in the deflection pattern, but instead they saw no change in the deflection pattern in accord with NPRF. We next review what this ``average-only'' projection per NPRF + $h$ (or more generally, per Information Invariance \& Continuity) tells us about entanglement via the Bell states. 

\newpage

\section{\label{SecBell}Implication for Entanglement via the Bell States}

Since the qubit forms the foundation of all (finite) denumerable-dimensional QM built in composite fashion, the most fundamental entangled states are the Bell states given by
\begin{eqnarray}
&|\psi_-\rangle = \frac{|u\rangle\otimes|d\rangle \,- |d\rangle\otimes|u\rangle}{\sqrt{2}} \nonumber \\
&|\psi_+\rangle = \frac{|u\rangle\otimes|d\rangle + |d\rangle\otimes|u\rangle}{\sqrt{2}} \label{BellStates} \\
&|\phi_-\rangle = \frac{|u\rangle\otimes|u\rangle \,- |d\rangle\otimes|d\rangle}{\sqrt{2}} \nonumber \\
&|\phi_+\rangle = \frac{|u\rangle\otimes|u\rangle + |d\rangle\otimes|d\rangle}{\sqrt{2}} \nonumber 
\end{eqnarray}
in the $\sigma_z$ eigenbasis. These correspond to the following density matrices
\begin{eqnarray}
&\rho_{\mbox{\tiny Sing}} = |\psi_-\rangle\langle\psi_-| = \left(I - \sigma_x\otimes\sigma_x - \sigma_y\otimes\sigma_y - \sigma_z\otimes\sigma_z    \right)/4 \nonumber \\
&\rho_{\mbox{\tiny TripZ}} = |\psi_+\rangle\langle\psi_+| = \left(I + \sigma_x\otimes\sigma_x + \sigma_y\otimes\sigma_y - \sigma_z\otimes\sigma_z  \right)/4 \label{BellStatesRho} \\ 
&\rho_{\mbox{\tiny TripX}} = |\phi_-\rangle\langle\phi_-| = \left(I - \sigma_x\otimes\sigma_x + \sigma_y\otimes\sigma_y + \sigma_z\otimes\sigma_z  \right)/4 \nonumber \\
&\rho_{\mbox{\tiny TripY}} = |\phi_+\rangle\langle\phi_+| = \left(I + \sigma_x\otimes\sigma_x - \sigma_y\otimes\sigma_y + \sigma_z\otimes\sigma_z  \right)/4  \nonumber 
\end{eqnarray}
As always, Alice and Bob are making their measurements on each of the two Bell state particles. If Alice makes her spin measurement $\sigma_1$ with her SG magnets oriented in the $\hat{a}$ direction and Bob makes his spin measurement $\sigma_2$ with his SG magnets oriented in the $\hat{b}$ direction, then
\begin{eqnarray}
    &\sigma_1 = \hat{a}\cdot\vec{\sigma}=a_x\sigma_x + a_y\sigma_y + a_z\sigma_z \nonumber \\
    &\sigma_2 = \hat{b}\cdot\vec{\sigma}=b_x\sigma_x + b_y\sigma_y + b_z\sigma_z  \label{sigmas}
\end{eqnarray}

The first state $|\psi_-\rangle$ is called the ``singlet state'' and it is invariant under any of the SU(2) transformations $e^{\textbf{i}\Theta\sigma_x}$, $e^{\textbf{i}\Theta\sigma_y}$, or $e^{\textbf{i}\Theta\sigma_z}$, corresponding to rotations of the SG magnets about the $x$, $y$, or $z$ axes, respectively (for computational details applicable to this section, see \cite{stuckey2020,silberstein2021}). This fact aligns with the signature of $\rho_{\mbox{\tiny Sing}}$ in Eq. (\ref{BellStatesRho}). $|\psi_-\rangle$ represents a total conserved spin angular momentum of zero ($S = 0$) for the two particles involved, i.e., Alice and Bob always obtain opposite outcomes ($ud$ or $du$) when making the same measurement. The other three states are called the ``triplet states'' and they are invariant under the SU(2) transformations $e^{\textbf{i}\Theta\sigma_z}$, $e^{\textbf{i}\Theta\sigma_x}$, and $e^{\textbf{i}\Theta\sigma_y}$, respectively. Again, note the correspondence with the signatures in Eq. (\ref{BellStatesRho}), respectively. They represent a total conserved spin angular momentum of one ($S = 1$, in units of $\hbar = 1$) in each of the spatial planes $xy$ ($|\psi_+\rangle$), $yz$ ($|\phi_-\rangle$), and $xz$ ($|\phi_+\rangle$). [To see this for $|\psi_+\rangle$, you have to transform the state to either the $\sigma_x$ or $\sigma_y$ eigenbasis where it has the same form as $|\phi_-\rangle$ or $|\phi_+\rangle$, respectively\cite{stuckey2020}.] Thus, Alice and Bob always obtain the same outcomes ($uu$ or $dd$) when measuring at the same angle in the symmetry plane of the relevant triplet state, i.e., when they share the same reference frame. In all four cases, the entanglement represents the conservation of spin angular momentum for the process creating the state. 

Suppose Alice and Bob are making measurements on their particles in the symmetry plane of a triplet state such that $\hat{a}\cdot\hat{b} = \cos{(\theta})$. Partition the data according to Alice's equivalence relation (her $\pm 1$ outcomes) and look at her $+1$ outcomes. Since we know Bob would have also measured $+1$ if $\theta$ had been zero (i.e., if Bob was in the same reference frame), we have exactly the same classical expectation depicted in Figures \ref{SGExp2} \& \ref{Projection} for the single qubit measurement, i.e., 
\begin{equation}
2P(+1,+1\mid \theta)(+1) + 2P(+1,-1\mid \theta)(-1) = \cos (\theta) \label{BA+}
\end{equation}
NPRF also requires that Alice and Bob each observe $+1$ half of the time and $-1$ half of the time, and that $P(-1,+1\mid \theta) = P(+1,-1\mid \theta)$, so we have  
\begin{equation}
\begin{aligned}
P(+1,+1\mid\theta) + P(+1,-1\mid \theta) & = \frac 12 \\
P(+1,-1\mid\theta) + P(-1,-1\mid \theta) & = \frac 12
\end{aligned}
\end{equation}
We can now solve these for the joint probabilities
\begin{equation}
P(+1,+1 \mid \theta) = P(-1,-1 \mid \theta) = \frac{1}{2} \mbox{cos}^2 \left(\frac{\theta}{2} \right) \label{QMjointLike}
\end{equation}
and 
\begin{equation}
P(+1,-1 \mid \theta) = P(-1,+1 \mid \theta) = \frac{1}{2} \mbox{sin}^2 \left(\frac{\theta}{2} \right) \label{QMjointUnlike}
\end{equation}
in accord with QM.

In other words, while spin angular momentum is conserved exactly when Alice and Bob are making measurements in the same reference frame, it is conserved only \textit{on average} when they are making measurements in different reference frames (related by SO(3) as shown in Figure \ref{ComplBases}). This ``surprising'' result is a direct consequence of NPRF + $h$, exactly as length contraction and time dilation are a direct consequence of NPRF + $c$. And, we could also partition the data according to Bob's equivalence relation (his $\pm 1$ results), so that it is Bob who claims Alice must average her results to satisfy ``average-only'' conservation. This is totally analogous to the relativity of simultaneity in SR. There, Alice partitions spacetime per her equivalence relation (her surfaces of simultaneity) and says Bob's meter sticks are short and his clocks run slow, while Bob can say the same thing about Alice's meter sticks and clocks per his surfaces of simultaneity. 

Of course, there is nothing unique about SG spin measurements except they can be considered direct measurements of $h$. The more general instantiation of NPRF per Information Invariance \& Continuity described in Section \ref{SecQubit} applies to \textit{any} qubit. So, for example, if we are again talking about photons passing or not passing through a polarizing filter, we would have ``average-only'' transmission for photons instead of ``average-only'' projection for spin-$\frac{1}{2}$ particles, both of which give ``average-only'' conservation of spin angular momentum between the reference frames of different mutually complementary measurements. So, most generally, the information-theoretic principle of Information Invariance \& Continuity leads to ``average-only'' \rule{1cm}{0.15mm} (fill in the blank) giving ``average-only'' conservation of the measured quantity for its Bell states. 

We should point out that the trial-by-trial outcomes for this ``average-only'' conservation can deviate substantially from the target value required for explicit conservation per Alice or Bob's reference frame. For example, we might have Bob's $+1$ and $-1$ outcomes averaging to zero as required for the conservation of spin angular momentum per Alice's reference frame. Thus, Alice says Bob's measurement outcomes are violating the conservation of spin angular momentum as egregiously as possible on a trial-by-trial basis. However, from the perspective of Bob's reference frame, it is Alice's  $+1$ and $-1$ outcomes averaging to zero that violate the conservation of spin angular momentum as egregiously as possible on a trial-by-trial basis. In classical physics, our conservation laws hold on average because they hold explicitly for each and every trial of the experiment (within experimental limits). But here, that would require a preferred reference frame. Thus, ``average-only'' conservation distinguishes classical mechanics and QM just as the relativity of simultaneity distinguishes Newtonian mechanics and SR. Consequently, we see that ``average-only'' conservation does not \textit{resolve} the mystery of quantum entanglement, it \textit{is} the mystery, i.e., it is what needs to be explained. 

What we've seen here is that we can explain ``average-only'' conservation as conservation that results necessarily from Information Invariance \& Continuity, which is conservation per NPRF in spacetime, precisely as the relativity of simultaneity results necessarily from NPRF + $c$. Of course, this then explains why quantum joint probabilities for the Bell states violate the Clauser-Horne-Shimony-Holt (CHSH) inequality precisely to the Tsirelson bound \cite{cirelson1980,landau1987,khalfin1992,stuckey2019}. That fact obtains because the quantum joint probabilities for the Bell states are precisely those that satisfy conservation in accord with NPRF. In contrast, classical probability theory would satisfy the CHSH inequality by requiring a preferred reference frame, thereby violating the invariant measurement of a fundamental constant of Nature $h$. Thus, the reconstructions of QM reveal the relativity principle at the foundation of QM precisely as it exists at the foundation of SR, i.e., demanding the invariant measurement of a fundamental constant of Nature. We now show how this bears on the no-signalling, ``superquantum'' joint probabilities of Popescu \& Rohrlich.

\newpage

\section{Ruling Out Superquantum PR-Box Probabilities}\label{SecPRbox}

In this section, we apply conservation per NPRF in spacetime (implied by conservation per Information Invariance \& Continuity) to answer an important question in quantum information theory, i.e., why don't we find any so-called ``superquantum'' joint probabilities in Nature? That is, assuming only that joint probabilities cannot permit faster-than-light signalling, Popescu \& Rohrlich introduced these ``no-signalling'' joint probabilities (the ``PR-box'') \cite{PR1994} that violate the CHSH inequality beyond the Tsirelson bound of QM
\begin{eqnarray}
&P(+1,+1 \mid a\phantom{^\prime},b\phantom{^\prime}) = P(-1,-1 \mid a\phantom{^\prime}, b\phantom{^\prime})=\frac{1}{2} \nonumber \\
&P(+1,+1 \mid a\phantom{^\prime},b^\prime) = P(-1,-1 \mid a\phantom{^\prime}, b^\prime)=\frac{1}{2} \label{PRcorr} \\
&P(+1,+1 \mid a^\prime,b\phantom{^\prime}) = P(-1,-1 \mid a^\prime, b\phantom{^\prime})=\frac{1}{2} \nonumber \\
&P(+1,-1 \mid a^\prime,b^\prime) = P(-1,+1 \mid a^\prime, b^\prime)=\frac{1}{2} \nonumber 
\end{eqnarray}
where each of $a^{\prime},a,b^{\prime}$, or $b$ is a measurement setting (H or T in what follows). ``No-signalling'' means that, for a given experimental setting, one observer's outcome is independent of the other observer's setting. For example, suppose that Alice chooses setting $a$. Then no-signalling implies that her observed outcome ($+1$ or $-1$) is independent of whether Bob chooses setting $b$ or $b'$. Specifically, the probability that Alice observes $+1$ given that she chooses setting $a$ and Bob chooses setting $b$ is $P(+1,+1|a,b) + P(+1,-1|a,b) = 1/2$ and this must equal the probability that she observes $+1$ if Bob chooses $b'$ instead, $P(+1,+1|a,b') + P(+1,-1|a,b') =1/2$.  Quantum information theorists quickly realized that if the joint probabilities in Eq. (\ref{PRcorr}) could be instantiated physically, they would provide an ``unreasonably effective'' means of communication. Let us follow Bub \& Bub's example \cite{bub2018} of a ``quantum guessing game'' to illustrate this point.

To help visualize the PR-box, the Bubs introduce superquantum coins called ``quoins.'' If both members of the pair of entangled quoins are flipped starting as heads (HH), they will end up tails-heads ($-1,1$) or heads-tails ($1,-1)$, never heads-heads ($1,1$) or tails-tails ($-1,-1$) (``$\ne$'' in Table \ref{tab:QuoinMechanics}). This corresponds to the last PR-box probability with $a^\prime = b^\prime$ = H. Given any other starting configuration, i.e., tails-tails (TT) or tails-heads (TH) or heads-tails (HT), they will end up ``equal,'' i.e., heads-heads ($1,1$) or tails-tails ($-1,-1$) (Table \ref{tab:QuoinMechanics}). These are magical coins indeed and as the Bubs show, their behavior cannot be explained by any ``rigging'' based on their starting positions. [Keep in mind at this point we are simply trying to understand how each quoin might behave deterministically to explain the Quoin Mechanics of Table \ref{tab:QuoinMechanics}, not the probabilistic PR-box itself.]

\newpage

\begin{table}
    \centering
    \begin{tabular}{cc|ccc|c}
    \multicolumn{2}{c}{} & \multicolumn{3}{c}{Alice's Quoin} \\
    &&  & { \textbf{H}} & \textbf{T} \\
    \cline{2-6}
  Bob's &  { \textbf{H}} & { } &  { $\ne$} & $=$ & \\
     Quoin & { \textbf{T}} & { } &  { $=$} & $=$ & \\
       \cline{2-6}
     
    \end{tabular}
    \caption{\textbf{Quoin Mechanics.} Results of flipping pairs of entangled quoins in four possible starting combinations.}
    \label{tab:QuoinMechanics}
\end{table}

To show this, they note there are only four ways to ``rig'' a quoin:
\begin{enumerate}
\item No matter how it starts, it ends up heads (H).
\item No matter how it starts, it ends up tails (T).
\item It stays the same way it starts (S for same).
\item It changes from the way it starts (O for other).
\end{enumerate}
They then list all the possible riggings for a heads-heads (HH) start that yield non-equal outcomes and all the possible riggings for a heads-tails (HT) or tails-heads (TH) or a tails-tails (TT) start that yield equal outcomes. As a result, it is apparent that there is no rigging that can yield the Quoin Mechanics of Table \ref{tab:QuoinMechanics}. Therefore, we simply have to accept Quoin Mechanics without knowing how they are instantiated in order to explore their implications for the Bubs' quantum guessing game.

Note, Mermin showed something similar in trying to explain quantum joint probabilities for the spin triplet state using ``instruction sets'' \cite{mermin1981,ross2020,stuckey2020}. In that case for quantum probabilities, as with these superquantum probabilities, the instruction sets didn't work and we were left with no ``constructive account'' of those quantum probabilities. Of course, the difference is that while we don't have any (consensus) ``causal mechanism'' to explain the quantum probabilities for the Bell states, we do instantiate them in the lab. And, as we just showed in Section \ref{SecBell}, those quantum probabilities do obey a very reasonable conservation principle (conservation per NPRF) while we will see that the PR-box probabilities violate that principle. Now we show how these conservation-violating PR-box probabilities are ``miraculous'' in their information exchanging capability. We start with the game itself.

\begin{table}
    \centering
    \begin{tabular}{|c|c|c|c|c|c|}
    \hline
    {\bf Lane} & \hspace*{0.2in} {\bf 1} \hspace*{0.2in} &  \hspace*{0.2in} {\bf 2}  \hspace*{0.2in} &  \hspace*{0.2in} {\bf 3}  \hspace*{0.2in} &  \hspace*{0.2in} {\bf 4}  \hspace*{0.2in} &  \hspace*{0.2in} {\bf 5}  \hspace*{0.2in} \\
    \hline
    \hline
    Bob's values & 1 & 0 & 1 & 1 & 0 \\
    \hline
    \multicolumn{6}{|c|}{\cellcolor{gray!25} BARRIER} \\
    \hline
     Alice's values & 1 & 0 & 0 & 1 & 1 \\
     \hline
    \end{tabular}
    \caption{\textbf{An Example Game Table.}  When Alice and Bob flip their quoins for their 1 and 0 values shown here for each lane, together they will have an even number of H in lanes 1 and 4 and an odd number of H in lanes 2, 3, and 5 per Quoin Mechanics (Table \ref{tab:QuoinMechanics}).}
    \label{tab:gametable}
\end{table}

\newpage

As you can see in Table \ref{tab:gametable}, there are 5 lanes on each side of the barrier, which does not allow Alice to see Bob's values and vice-versa. The lanes are numbered 1 through 5 and the game is started by the dealer setting values of 1 or 0 for each player in each of the five lanes (Table \ref{tab:gametable}). After the dealer has set these ten values, we see that there are some lanes that have a 1 on both sides of the barrier (lanes 1 and 4 in Table \ref{tab:gametable}). Alice is the ``guesser'' and because of the barrier she doesn't know what Bob has in his lanes. Her job is to guess whether there are an even number of lanes that have a 1 on both sides (as is the case in Table \ref{tab:gametable}) or an odd number of such lanes. The only way she can know the answer with certainty is if the dealer were to set all five of her lanes to 0 in which case she knows the answer is even, i.e., there are zero lanes that have a 1 on both sides. Of course, the dealer isn't going to do that, so she will always have at least one 1 on her side. For all such cases, it is not difficult to convince yourself that the probability of the right answer being odd(even) is 50\%(50\%). Alice and Bob buy six poker chips to play and the House doubles their chips for a correct guess. [They turn their poker chips back into money when they leave the Quasino of course.] If Alice guesses wrong, they lose all their chips to the House. Since the odds are 50-50 of guessing right, Alice and Bob end up breaking even if they continue to play, meaning they'll not win or lose money overall on average.

But, there is another aspect of the game that we have not told you -- Bob can spend one poker chip to send one bit of information to Alice. In that case, Alice and Bob have five chips remaining and if they win, the House pays them five chips, i.e., the House doubles their remaining poker chips. Likewise, Bob can spend two chips to send two bits of information to Alice which means the House would pay four chips if they win. Once they have to spend three chips to win, the House pays three chips and they break even monetarily by winning the game. For example, Alice could ask Bob to send her the values in his lanes 1, 4, and 5 of Table \ref{tab:gametable}, so she can see if he has any 1's in those lanes. That's all the information she needs, since she has 0's in lanes 2 and 3. Bob sends the bits 1, 1, and 0 which tell her they have just two lanes (lanes 1 and 4) where they both have a 1, so the answer is ``even.'' They win the game and three chips, but they spent three chips to send the three bits of information, so they broke even monetarily. You can easily convince yourself that this strategy would actually end up losing money in the long run.

Now it's time to show that if Alice and Bob use quoins, then they can win this game every time by merely passing one bit of information each game, thereby winning five chips every game, i.e., netting four chips every game. Here is the Bubs' strategy. Alice and Bob start with five pairs of quoins. [Don't confuse these with the six chips they bought from the House to play the game, they brought these five pairs of quoins with them to the Quasino.] Again, Alice is the guesser, so it is Bob who will be sending the one bit of information. Alice's quoins are labeled 1A, 2A, 3A, 4A, 5A and Bob's are labeled 1B, 2B, 3B, 4B, 5B. Quoins 1A and 1B are entangled per Quoin Mechanics, as are 2A and 2B, etc. The number on each quoin corresponds to each lane of the game. For all lanes in which Alice has a 1, she flips the corresponding quoin starting with H. In all lanes with a 0, she flips the corresponding quoin starting with T. Bob does likewise with his quoins for his values and lanes. Bob then pays one chip to send Alice one bit of information, i.e., 1 if he has an odd number of H's and 0 if he has an even number of H's. From this one bit of information, Alice now knows with certainty whether they have an even or odd number of lanes with two 1's. The strategy is simple although the reasoning behind it is not trivial.

The key is to observe that the individual outcomes of Alice and Bob's quoin tosses do not really matter, but their pairs do matter. In any lane with two 1's, Alice and Bob together observe a total of one H after they flip their quoins for that lane (Quoin Mechanics), which is odd. In the other lanes, they observe an even number (0 or 2) of H's (Quoin Mechanics). Therefore, their \textit{combined} count of H's is even if and only if there are an even number of lanes with two 1's. Neither player by themselves knows whether the combined number of H's is even or odd because each person can only see the outcomes of his/her own quoins. But, all Alice needs to know after flipping her five quoins is whether or not Bob has an even number of H's or an odd number of H's, and he can send her that one bit of information (1 for odd and 0 for even, for example) by spending just one chip. Thus, Quoin Mechanics guarantees they will win five chips every game (netting four).

How egregious is this advantage? In other words, can quantum joint probabilities achieve anywhere near this success rate? Let's look at the PR-box. We see that the fourth PR-box probability corresponds to the HH case producing unequal outcomes in Quoin Mechanics. That is, using $a = b$ = T and $a^\prime = b^\prime$ = H with outcomes +1 = H and --1 = T, the PR-box aligns with Quoin Mechanics (Table \ref{tab:QuoinMechanics}). Since it is the HH case that allows us to discern even or odd pairs of 1's in our guessing game using only one bit of information, let's scrutinize the fourth PR-box probability using our conservation principle.

According to QM, the joint probability of measuring like results for a triplet state in its symmetry plane is $\mbox{cos}^2\left(\frac{\theta}{2}\right)$ and the joint probability of measuring unlike results is $\mbox{sin}^2\left(\frac{\theta}{2}\right)$, where $\theta$ is the angle between $\hat{a}$ and $\hat{b}$. The first PR-box probability says that $\hat{a} = \hat{b}$ (same results), the second PR-box probability says $\hat{a} = \hat{b}^\prime$ (same results), and the third PR-box probability says $\hat{a}^\prime = \hat{b}$ (same results). Thus, these three PR-box joint probabilities in total say $\hat{a} = \hat{a}^\prime = \hat{b} = \hat{b}^\prime$. So, we need the fourth PR-box probability to say $\hat{a}^\prime = \hat{b}^\prime$, i.e., same results, but of course it says we must get opposite results, which means $\hat{a}^\prime = -\hat{b}^\prime$. Therefore, the PR-box joint probabilities violate our conservation principle in a maximal fashion for the triplet states. [A similar argument can be made using the singlet state.]

Consequently, in order to satisfy our conservation principle we need the outcomes of a HH start to be equal just as the outcomes of a TH, HT, or TT start. But, if we make the outcomes of a HH start equal, we lose the advantage of Quoin Mechanics. In fact, using quantum coins (the outcomes of a HH start are equal) instead of superquantum quoins (the outcomes of a HH start are not equal) puts us right back to a 50-50 chance of winning the guessing game (recall the reasoning behind the quoin strategy). So, we see that the superquantum PR-box probabilities are not just a little bit better than quantum probabilities for the quantum guessing game, they are ``unreasonably effective.'' But, again, they violate our conservation principle, so it is probably the case that a physical instantiation of the PR-box probabilities is a pipe dream akin to a perpetual motion machine.

\section{\label{SecConcl}Conclusion}

Quantum information theorists have a produced a principle account of denumerable-dimensional QM whereby ``quantum reality'' is characterized most succinctly by the information-theoretic principle of Information Invariance \& Continuity. Accordingly, reality is composed fundamentally of discrete bits of irreducible, finite information that can be instantiated and measured physically in 3-dimensional space by measurement devices which are themselves composed of such bits (``closeness requirement''). For the Bloch sphere, Information Invariance \& Continuity reflects the fact that it is always possible to create a path from one pure state to another by passing through pure states only, i.e., the surface of the Bloch sphere is composed of pure states. This is quite a ``surprising'' fact from the point of view of classical probability theory where the only path in probability state space between pure states is through mixed states with lower information content. The higher-dimensional and multi-particle Hilbert space structure of QM can all be built from this fundamental, ``surprising'' qubit structure characterized by Information Invariance \& Continuity. 

To help clarify the significance of this information-theoretic result, we applied it to the Stern-Gerlach (SG) measurements of a spin-$\frac{1}{2}$ particle. In that case, the Pauli matrices are used to represent spin-$\frac{1}{2}$ measurements $J_i = \frac{\hbar}{2}\sigma_i$ with $[J_x,J_y]=\textbf{i}\hbar J_z$, cyclic, which are responsible for the reference frames associated with complete sets of mutually complementary spin measurements. Information Invariance \& Continuity means the spin measurement operators are related by SU(2) in Hilbert space, which means the corresponding reference frames are related by SO(3) in real space. Thus, invariance of the eigenvalues under SU(2) means invariance of measurement outcomes in real space under SO(3), which is a transformation subgroup of both the Lorentz and Galilean transformation groups between inertial reference frames. Since SG spin measurements constitute a measurement of Planck's constant $h$, Information Invariance \& Continuity entails NPRF + $h$ in exact analogy to NPRF giving rise to the invariant measurement of the speed of light $c$ at the foundation of SR. Of course, it must be the case that NPRF + $h$ is responsible for the non-commutative algebra of the spin measurement operators to begin with.

\newpage

To see that, suppose $|\psi\rangle = |u\rangle$ along $\hat{z}$. Then classically, we know the exact measurement outcome along $\hat{b}$ will be $\hat{b}\cdot\hat{z} = \cos{\theta}$. In other words, the measurement outcome in one reference frame ($+1$ in the $\hat{z}$ frame) determines the exact measurement outcome in another reference frame ($\cos{\theta}$ in the $\hat{b}$ frame) in the classical case. However, this violates NPRF because there is only one reference frame with the ``right'' eigenvalue ($+1$ in the $\hat{z}$ frame) and therefore only one frame that measures the correct value for $h$. So in the classical case, the different measurement operators commute, e.g., $[J_x,J_y] = 0$. In contrast, NPRF (and Information Invariance \& Continuity) says we must obtain $\pm 1$ along $\hat{b}$ just like any other direction. Thus, NPRF does not allow us to deduce the exact measurement outcome in the $\hat{b}$ reference frame using our $+1$ measurement outcome in the $\hat{z}$ reference frame. Again, this is what Brukner \& Zeilinger meant when they said the qubit does not contain enough information to account for the outcomes of every possible measurement done on it, so a theory of qubits must be probabilistic \cite{zeilinger1999,bruknergroup,bruknerZeil1999}. Thus in the quantum case, the different measurement operators do not commute, i.e., $[J_x,J_y] = i\hbar J_z$. Again, just like elsewhere in QM, letting Planck’s constant go to zero recovers classical physics. Here we see that NPRF + $h$ is responsible for the non-commutative algebraic structure of QM, in contrast to the commutative algebraic structure of classical mechanics.

Since NPRF (and Information Invariance \& Continuity) requires we obtain $\pm 1$ along $\hat{b}$ just like any other direction, the classically expected result of $\hat{b}\cdot\hat{z} = \cos{\theta}$ can obtain only \textit{on average}. Thus, NPRF + $h$ gives us ``average-only'' projection for measurement outcomes in different reference frames. When applied to Bell state entanglement, we showed that ``average-only'' projection for one particle leads to ``average-only'' conservation between two entangled particles in different reference frames. Thus, according to the reconstructions of QM the mysterious ``average-only'' conservation of Bell state entanglement, and therefore the Tsirelson bound, follow from conservation per Information Invariance \& Continuity, which is conservation per NPRF in spacetime. Finally, we showed that a hypothetical instantiation of the superquantum PR-box joint probabilities in spacetime violates conservation per the relativity principle, thereby allowing one to communicate with ``unreasonable'' effectiveness per Bub \& Bub \cite{bub2018}. Thus, it seems unlikely that the PR-box can be realized in Nature, even though these joint probabilities do not violate the no-signalling condition. How conservation per Information Invariance \& Continuity bears on other information-theoretic phenomena, e.g., macroscopic entanglement witnesses \cite{terhal2000,entanglementWitness2014,bruknerEntangleWitness}, is left for future work.

\newpage

Whether or not one believes principle accounts are explanatory is irrelevant here. No one disputes what the postulates of SR are telling us about Nature, even though there is still today no (consensus) constructive account of time dilation and length contraction, i.e., there is no ``interpretation'' of SR. While Lorentz complained \cite[p. 230]{lorentz}:
\begin{quote}
Einstein simply postulates what we have deduced, with some difficulty and not altogether satisfactorily, from the fundamental equations of the electromagnetic field.
\end{quote} 
\noindent{he nonetheless acknowledged \cite[p. 230]{lorentz}}:
\begin{quote}
By doing so, [Einstein] may certainly take credit for making us see in the negative result of experiments like those of Michelson, Rayleigh, and Brace, not a fortuitous compensation of opposing effects but the manifestation of a general and fundamental principle.
\end{quote} 
We have shown how the principle of Information Invariance \& Continuity at the basis of axiomatic reconstructions of QM provides an understanding of the qubit and Bell state entanglement that is every bit the equal of Einstein's postulates of SR for understanding time dilation and length contraction. Thus, it is no longer true that ``nobody understands quantum mechanics'' unless it is also true that nobody understands special relativity. Very few physicists would make that claim.

\section*{References}
\raggedright
\bibliographystyle{iopart-num}
\bibliography{biblio}
\end{document}